\documentclass[twocolumn,trackchanges,twocolappendix]{aastex631}
\usepackage{natbib}
\usepackage{color}
\usepackage{graphicx}
\usepackage{multirow}

\graphicspath{{./}{plots/}}

\accepted{December 11, 2024}
\submitjournal{ApJ}

\shorttitle{Toward early-type systems as extragalactic milestones. IV}
\shortauthors{Taormina et al.}

\begin{document}

\title{Toward early-type eclipsing binaries as extragalactic milestones: IV. Physical properties of three detached O/B-type systems in the LMC}

\correspondingauthor{M{\'o}nica Taormina}
\email{taormina@camk.edu.pl}

\author[0000-0002-1560-8620]{M{\'o}nica Taormina}
\affiliation{Centrum Astronomiczne im. Miko{\l}aja Kopernika PAN, Bartycka 18, 00-716 Warsaw, Poland}

\author[0000-0002-9443-4138]{G. Pietrzy{\'n}ski}
\affiliation{Centrum Astronomiczne im. Miko{\l}aja Kopernika PAN, Bartycka 18, 00-716 Warsaw, Poland}

\author{R.-P.  Kudritzki}
\affiliation{Institute for Astronomy, University of Hawaii at Manoa, Honolulu, HI 96822, USA}
\affiliation{Munich University Observatory, Munich, Germany}

\author[0000-0003-3861-8124]{B. Pilecki}
\affiliation{Centrum Astronomiczne im. Miko{\l}aja Kopernika PAN, Bartycka 18, 00-716 Warsaw, Poland}

\author[0000-0001-5130-6040]{I. B. Thompson}
\affiliation{Carnegie Observatories, 813 Santa Barbara Street, Pasadena, CA 91101-1292,  USA}

\author[0000-0002-0874-1669]{J. Puls}
\affiliation{LMU M\"unchen, Universit\"atssternwarte, Scheinerstr. 1, 81679 M\"unchen, Germany}

\author[0000-0002-3125-9088]{M. G\'{o}rski}
\affiliation{Centrum Astronomiczne im. Miko{\l}aja Kopernika PAN, Bartycka 18, 00-716 Warsaw, Poland}

\author[0000-0003-1515-6107]{B. Zgirski}
\affiliation{Universidad de Concepci\'{o}n, Departamento de Astronom\'{i}a, Casilla 160-C, Concepci\'{o}n, Chile} 

\author[0000-0002-7355-9775]{D. Graczyk}
\affiliation{Centrum Astronomiczne im.  Miko{\l}aja Kopernika PAN, Rabia{\'n}ska 8, 87-100 Toru{\'n}, Poland}

\author[0000-0002-9424-0501]{M. A. Urbaneja}
\affiliation{Universit\"at Innsbruck, Institut f\"ur Astro- und Teilchenphysik, Technikerstr. 25/8, 6020 Innsbruck, Austria}

\author[0000-0003-1405-9954]{W. Gieren}
\affiliation{Universidad de Concepci\'{o}n, Departamento de Astronom\'{i}a, Casilla 160-C, Concepci\'{o}n, Chile}

\author[0000-0003-2335-2060]{W. Narloch}
\affiliation{Centrum Astronomiczne im. Miko{\l}aja Kopernika PAN, Bartycka 18, 00-716 Warsaw, Poland}

\author[0000-0003-0594-9138]{G. Hajdu}
\affiliation{Centrum Astronomiczne im. Miko{\l}aja Kopernika PAN, Bartycka 18, 00-716 Warsaw, Poland}

\begin{abstract}
We present a complete set of physical parameters for three early-type eclipsing binary systems in the Large Magellanic Cloud (LMC): OGLE LMC-ECL-17660, OGLE LMC-ECL-18794, and HV~2274, together with the orbital solutions. The first and third systems comprise B-type stars, while the second has O-type components and exhibits a total eclipse.
We performed a complex analysis that included modeling light and radial velocity curves, O-C analysis, and additional non-LTE spectroscopic analysis for the O-type system. We found that OGLE LMC-ECL-17660 is at least a triple and, tentatively, a quadruple. A significant non-linear period decrease was determined for HV 2274. Its origin is unclear, possibly due to a faint, low-mass companion on a wide orbit.
The analyzed components have masses ranging from 11.7 M$_\odot$ to 22.1 M$_\odot$, radii from 7.0 R$_\odot$ to 14.2 R$_\odot$, and temperatures between 22500 K and 36000 K. For HV~2274, the precision of our masses and radii is about six times higher than in previous studies.
The position of the components of all six systems analyzed in this series on the mass-luminosity and mass-radius diagrams indicates they are evolutionarily advanced on the main sequence. Our sample contributes significantly to the knowledge of physical parameters of early-type stars in the mass range of 11 M$_\odot$ to 23 M$_\odot$. A new mass-luminosity relation for O and B-type stars in the LMC is provided. Additionally, we used the measured apsidal motion of the systems to compare the observational and theoretical internal structure constant.

\end{abstract}


\section{Introduction} \label{sec:intro}

The accurate mass and radii are fundamental parameters that characterize a star. They can, in general, be determined with sufficient accuracy (having errors $\leq2\%$) if the star is in an eclipsing binary system with both components visible in the spectra \citep{andersen:1991, torres:2010}\footnote{Except for the closest to us systems, such as $\alpha$ Cen \citep{kervella:2017}.}. In such a case, the analysis of the periodic changes in brightness of the binary can be combined with that of radial velocities (RVs) of both components, providing a complete description of the system and the component stars. 
Accurate and precise measurement of stellar radii is even more critical for stars used as distance indicators, regardless of the applied method (e.g., surface brightness-color relation or spectral energy distribution fitting). Only with such measurements can we determine reliable distances to other galaxies and improve our knowledge of the cosmic distance scale. Therefore, it is crucial to perform a detailed and careful analysis of extragalactic binary systems based on high-quality data.
Moreover, studying early-type eclipsing binaries has a strong impact on the development of stellar evolutionary models and contributes to understanding stellar populations in their host galaxies. 

In this series of papers (\citealt{taormina:2019}, henceforth Paper I; \citealt{taormina:2020}, Paper II; and \citealt{taormina:2024}, Paper III), we perform a detailed analysis of a sample of early-type detached eclipsing binaries in the Large Magellanic Cloud (LMC), with the aim of calibrating the blue part (V-K $<$ 0 mag) of the surface brightness-color relation (SBCR). We have analyzed three systems to date, providing accurate physical parameters for all their components.

In the present work, we continue with the analysis of three more LMC detached eclipsing double line spectroscopic binaries from our sample: OGLE LMC-ECL-17660 (henceforth BLMC-04), OGLE LMC-ECL-18794 (henceforth BLMC-05) and HV 2274 (OGLE ID: LMC-ECL-05764; henceforth BLMC-06). As described in \citet[henceforth Paper I]{taormina:2019}, the systems were selected from the Optical Gravitational Lensing Experiment (OGLE) catalogs of eclipsing binaries in the LMC \citep{graczyk:2011,pawlak:2016}, based on their brightness and shape of the light curves (the depth and width of the eclipses, and eclipse depth ratio). This was to ensure the highest quality of results and that both components could be seen in the spectra.

In Table~\ref{tab:info}, we present the most relevant information from the literature on these systems, which consists of their coordinates, orbital period, brightness at maximum in the $I_c$ and $V$ bands, $(V-I_c)$ color, and spectral type. Detailed information and a literature background (if any) for each system are given in the Appendix section~\ref{sec:app} of the paper.

\begin{deluxetable*}{lcccccccc} \label{tab:info}
    \tabletypesize{\footnotesize}
     \tablecaption{General information from the literature }
         \tablehead{
                \colhead{Our ID} & \colhead{OGLE ID} & \multicolumn2c{Coordinates} & \colhead{Orbital Period} & \multicolumn2c{Brightness}  & \colhead{Color} & \colhead{Spectral Type}  \\
                \cline{3-4} \cline{6-7}
                \colhead{  }     & \colhead{  }      & \dcolhead{\alpha_{2000}} & \dcolhead{\delta_{2000}} & \dcolhead{[days]}   & \dcolhead{I_c}  & \dcolhead{V}    & \dcolhead{(V-I_c)}  & \colhead{} 
                   }
           \startdata
                BLMC-04  & LMC-ECL-17660 & 05:30:22.13   &  -69:14:52.0 &  6.229090   & 14.429 & 14.266 & -0.163 & ... \\
                BLMC-05  & LMC-ECL-18794 & 05:32:47.74   &  -68:37:26.6 &  5.946846   & 13.397 & 13.194 & -0.203 & O9.5III+B0 \tablenotemark{a}      \\
                BLMC-06  & LMC-ECL-05764 & 05:02:40.77   &  -68:24:21.4 &  5.725912   & 14.371 & 14.231 & -0.140 & B1.5III+B1.5III \tablenotemark{b}  
            \enddata
            \tablenotetext{a}{\cite{evans:2015}} \tablenotetext{b}{\cite{guinan:1998a}}            
\end{deluxetable*} 

The paper is organized as follow: In section~\ref{sec:data}, we present the data used in this study. In section~\ref{sec:analysis}, we explain the general steps taken to derive the fundamental parameters of the systems and describe additional phenomena that had to be taken into account for each object.
In section~\ref{sec:results}, we show full solutions for the systems and in section~\ref{sec:discussion} we discuss the results.

\section{Observational Data}  \label{sec:data}
As in our previous works of the series, we made use of the long-timespan photometric data of the third and fourth phases of the OGLE survey \citep{udalski:2003,udalski:2015}. Based on availability, they were complemented with data from other surveys. For BLMC-04, the OGLE photometry was supplemented with observations from the second phase of the Expérience pour la Recherche d'Objets Sombres survey \citep[EROS-2;][]{tisserand:2007}, and for two systems, with observations from the Massive Compact Halo Objects survey \citep[MACHO;][]{alcock:1997}. For one star, BLMC-05, no data from the EROS or MACHO surveys were available. Additionally, for all the systems, we used the near-infrared $K_s$-band photometry from the VISTA Magellanic Clouds IR photometric survey \citep[VMC;][]{cioni:2011} together with IR photometry collected during our runs with the SOFI imaging camera on the 3.58m ESO New Technology Telescope \citep[NTT;][]{moorwood:1998} at the La Silla Observatory, Chile. More details about how the observations, reduction, and photometry were performed can be found in Paper III. 

Time-series photometry from the Gaia Data Release 3 \citep[Gaia DR3,][]{gaia3:2023} are available only for two of the systems. These data were included in the analysis to have a longer time baseline, which is crucial to improve the determination of, e.g., the apsidal motion and period change.

In Table~\ref{table:photomData}, we summarize the photometric data used in the analysis of the systems, showing the number of measurements for a given band and from a given survey. For BLMC-06, additional CCD photometry in B, V, and Ic bands is available from \citet[][henceforth W92]{watson:1992}. These data were the base for the earlier studies of this system by \citet[][hereafter G98a]{guinan:1998a} and  \citet[][hereafter R00]{ribas:2000}. Unfortunately, this photometry is only relative and cannot be used to obtain the absolute magnitudes needed to calibrate SBCR. Moreover, these data are not numerous and only slightly precede the MACHO observations.
For these reasons and to save computing time, we did not include them in the binary modeling.

\begin{deluxetable*}{lcc c ccccc c cc ccc}
    \tabletypesize{\footnotesize}
    \tablecaption{Number of photometric data points per survey} \label{table:photomData}
         \tablehead{
                \colhead{Our ID} & \multicolumn2c{OGLE-III} &&  \multicolumn2c{OGLE-IV} & \colhead{EROS-2} & \multicolumn2c{MACHO}  && \multicolumn2c{IR data}  && \multicolumn2c{Gaia DR3}  \\
                \cline{2-3} \cline{5-6} \cline{8-9} \cline{11-12}  \cline{14-15}
                \colhead{  }     & \dcolhead{I_c} & \dcolhead{V} && \dcolhead{I_c} & \dcolhead{V}   & \colhead{$R$$_{eros}$}  & \dcolhead{R}   & \dcolhead{V}  && \dcolhead{K_s (VMC)}  & \dcolhead{K_s (our)}  && \dcolhead{BP}  & \dcolhead{RP} 
                   }
           \startdata
                BLMC-04  &  215   &  22   &&  590  &  149   &  483    &  514 & 513  && 15  & 11 &&  37  & 35 \\
                BLMC-05  &  590   &  40   &&  587  &  ...   &  ...    &  ... & ...  && 17  & 14  &&  ... & ... \\
                BLMC-06  &  447   &  43   &&  667  &  190   &  ...    &  475 & 471  && 16  & 11  &&  42 &  42
           \enddata
\end{deluxetable*} 

Regarding spectroscopic data, we used the high-resolution optical echelle spectra acquired with UVES \citep{dekker:2000} on the Very Large Telescope (VLT) at Paranal Observatory and with the MIKE spectrograph \citep{bernstein:2003} on the Magellan Clay Telescope at Las Campanas Observatory. The setup of the instruments and reductions were the same as described in Paper I.  A total of 12 high-resolution spectra (8 UVES + 4 MIKE) were obtained for BLMC-04, 15 (9 UVES + 6 MIKE) for BLMC-05, and 12 (7 UVES + 5 MIKE) for BLMC-06. The typical signal-to-noise ratios near $4000$~\r{A}  are about 50-60 for the UVES, and 30-40 for the MIKE spectra.

\section{Analysis} \label{sec:analysis}
\subsection{Combined radial velocity and light curve analysis}
We used the same methodology as in previous papers from this series to characterize our early-type detached eclipsing binaries and their components. We performed standard binary modeling of light and radial velocity curves supplemented by additional methods whenever it helped to obtain a better model.

We started by deriving the orbital parameters, for which we had first to extract RVs from the collected spectra. The RV measurements were done using the broadening function (BF) technique \citep{rucinski:1992, rucinski:1999} implemented in the {\tt RaveSpan} code \citep{pilecki:2017} to the reduced one-dimensional spectra. 
As reference (template) spectra, we used spectra of slowly rotating standard stars from the ESO Data Archive, which have spectral types similar to those estimated for the components of our systems.
Once we had extracted the radial velocities, we modeled the RV curves, taking into account the orbital elements of the binary: period ($P$), orbital semi-amplitudes ($K_i$), the velocity of the system's center-of-mass ($\gamma$), the reference time ($T_0$), the orbital eccentricity ($e$), and the argument of periastron passage ($\omega$) of the primary. We fitted the model to the observed RV curve of each system component and obtained a preliminary orbital solution. 

\begin{figure}
    \includegraphics[width=1.0\linewidth]{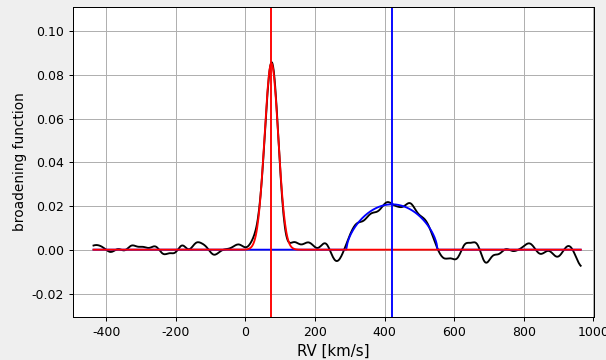} \\
    \includegraphics[width=1.0\linewidth]{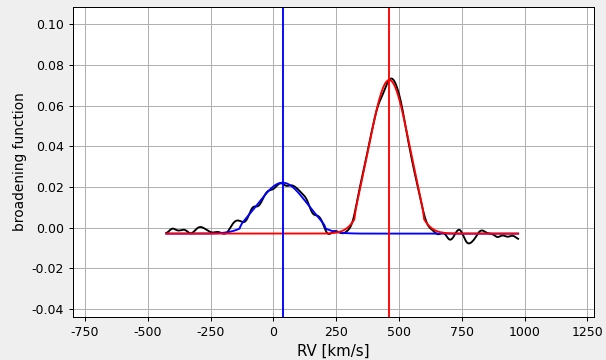} \\
    \includegraphics[width=1.0\linewidth]{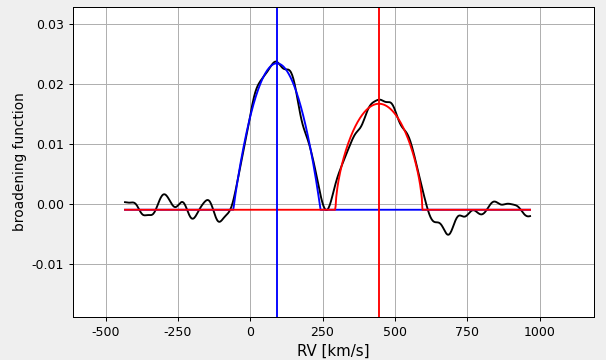} 
\caption{Example of the application of the BF method for the analyzed systems: BLMC-04 (top), BLMC-05 (middle), and BLMC-06 (bottom). The fitted profile of the primary component is shown in blue and of the secondary in red.  Wide profiles mean the component is a fast rotator, while a narrow peak means slow rotation. The surface under the profile is related to the total brightness of the component in the analyzed wavelength range.}
\label{rotation}
\end{figure}

In the BF method, the star's projected rotational velocity ($v_{prot} = v_{rot}\sin i)$\footnote{Here $i$ is the inclination of the rotation axis of the star, which in general does not have to be aligned with the orbital axis.} can be obtained directly from the fit of rotational profiles to the output broadening function. For each system, we determined the projected rotational velocities of both stars from several spectra with well-separated component lines, i.e., spectra taken close to the orbital quadratures. As such, we adopted the mean value of all individual determinations for each component. 
In Fig.~\ref{rotation}, we show an example of applying the BF method for all three analyzed systems. The area and width of the fitted profiles directly indicate which object is more luminous and which is a fast or slow rotator.
From this perspective, the most interesting system is BLMC-04 (top panel in Fig.~\ref{rotation}), where the primary component (blue fit) rotates relatively fast ($v_{prot}\sim$130 $km\,s^{-1}$), while the secondary (red fit) is a slow rotator ($v_{prot}\sim$25 $km\,s^{-1}$). Alternatively, the secondary component may have a significantly misaligned rotation axis, but this cannot be confirmed without additional time-consuming dedicated observations. The components of other systems have high and similar rotation velocities. For BLMC-05 these are $v_{prot}\sim$145 $km\,s^{-1}$ and 135 $km\,s^{-1}$ and for BLMC-06 it is about 150 $km\,s^{-1}$ for both stars. For our study, we assumed that the rotational axis is parallel to the orbital axis to calculate final rotational velocities ($v_{rot}$) from the observed projected values. For the secondary of BLMC-04 this assumption may be wrong, but even in that case, its effect on our results would be relatively minor. Further investigation is necessary for this system.
  
With this additional information from the spectroscopic data, we will be later able to compare the angular rotation of each component of the binary system with its orbital angular velocity. The ratio of these velocities, the synchronicity parameter $F=\omega_{rot}/\omega_{orb}$, is important for the light curve modeling, as it slightly changes the stars' shape. The rotation and orbital motions are synchronized if $F \approx 1$ (i.e., both velocities are similar).

In the next step, we modeled all the light curves simultaneously with the RV curves using a Python wrapper over the {\tt Phoebe 1} code \citep{prsa:2005}.
As only the temperature ratio is fitted in the modeling, we must independently determine a temperature for one of the components. In general, we do that for the more luminous of the two.
For the system with B spectral type components, BLMC-04, the secondary temperature of 28000 $\pm$ 1500 K was determined from the dereddened V-I and V-K colors using the color-temperature calibrations of \citet{worthey:2011}. For the other B-type system, BLMC-06, we used a secondary temperature of 23000 K $\pm$ 180 K derived by G98a from Hubble UV spectra. However, regarding its uncertainty, we adopted a more conservative value of 500 K (typical for our spectroscopic determinations for other systems) to consider a possible systematic difference between their and our methods and allow for a reliable comparison of the results.
As the aforementioned color-temperature relations do not work well for O-type stars and the literature does not provide a temperature determination for BLMC-05, we performed a non-LTE spectroscopic analysis to determine the temperature of its components for that system. This analysis is described in section~\ref{sec:non-lte}.

\begin{figure*}
    \begin{center}
        \includegraphics[width=0.90\textwidth]{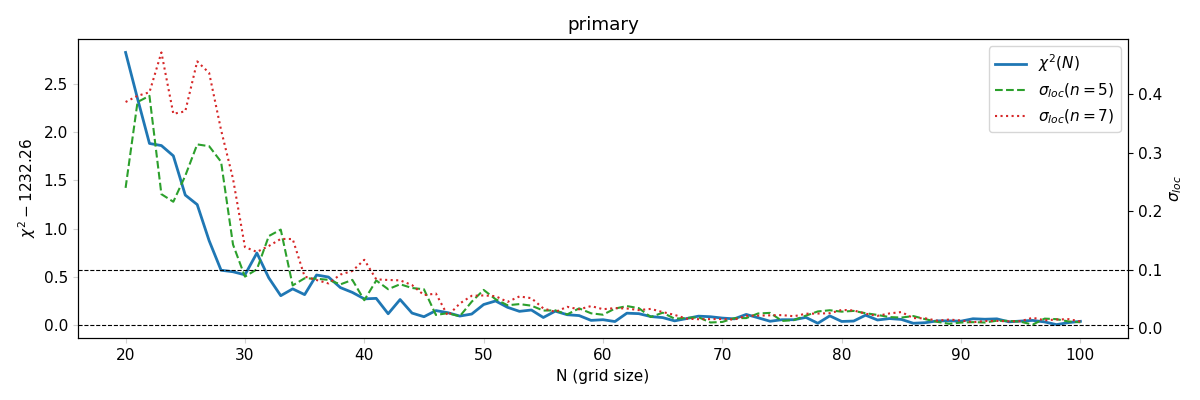} \\
        \includegraphics[width=0.90\textwidth]{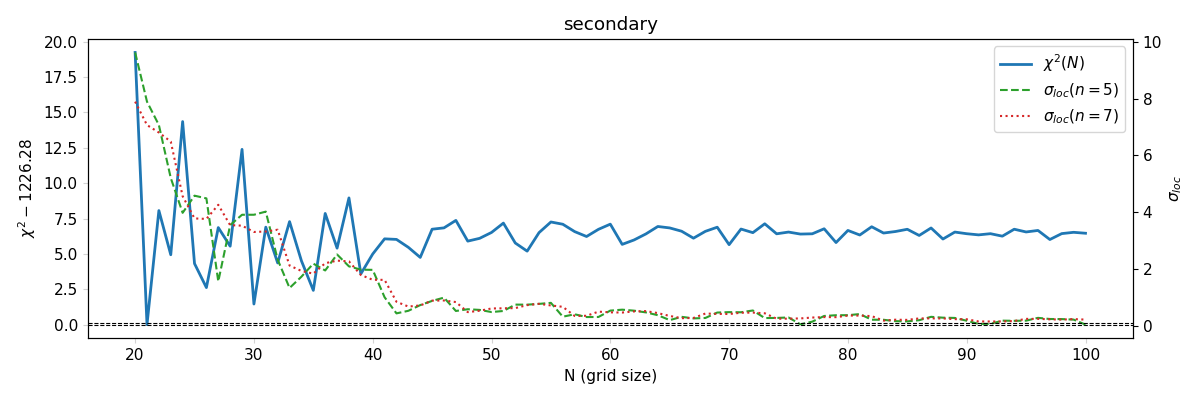} 
    \end{center}
    \caption{A plot showing the $\chi^2$ values depending on the grid size (N) used for the stellar surface discretization for a given component (solid blue line). The optimal values are selected according to the behavior of the local standard deviation ($\sigma_{loc}$) for n=5 and 7 nearest points along N. For the primary, the optimal N is 67, and for the secondary, it is 84. A horizontal line representing a variation on the $0.1-\sigma$ level is shown as a black dashed line for comparison.}
    \label{Ngrid}
\end{figure*}

In {\tt Phoebe 1}, it is necessary to specify the grid size (N) for the surface discretization, which highly influences the precision of the obtained model. It cannot be too low, but increasing it too much slows the code significantly. A problem of choosing an optimal value occurs, and eventually, it is often selected arbitrarily, the same for both components. For our analysis, we developed a method to choose an optimal grid size N for each component of a system individually, depending on its characteristics. Using a preliminary best fitting model, we calculate the $\chi^2$ value for a range of grid sizes between 20 and 100 and look at its variability, measured as local standard deviation ($\sigma_{loc}$) for n=5 or 7 nearest points, along the increasing N. As the optimal grid size for a given component, we choose N at which the $\chi^2$ stabilizes, i.e., $\sigma_{loc}$ converges at an approximately stable value close to zero (see Fig.~\ref{Ngrid}). Note that, in general, it is not the lowest $\chi^2$ value.
This way, we make sure that what we later find in the modeling as a minimum $\chi^2$ is not a numerical artifact, i.e., that related numerical variations in the $\chi^2$ value do not shift the parameter values by more than a small fraction of a $\sigma$. For the object in Fig.~\ref{Ngrid}, the values of N are 67 and 84 for the primary and secondary, respectively. We note here that the variation of $\chi^2$ for the secondary dominates the total $\chi^2$ variation, with $\sigma_{loc}$ stabilizing at a relatively high value compared to the primary. Generally, the grid size values we obtain this way exceed 65.

\begin{figure*}
    \begin{center}
      \includegraphics[width=0.32\textwidth]{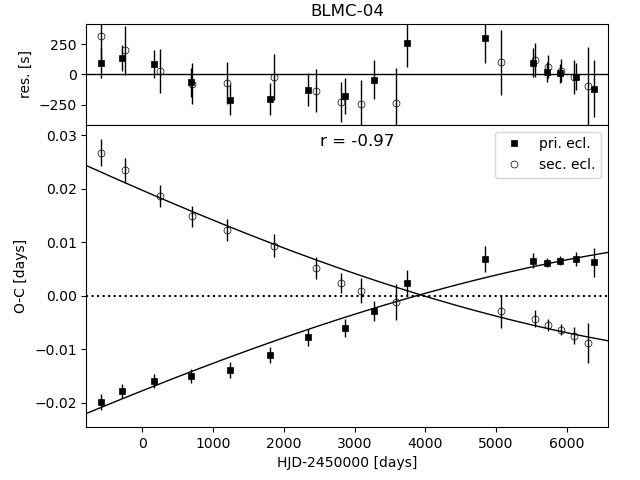}
      \includegraphics[width=0.32\textwidth]{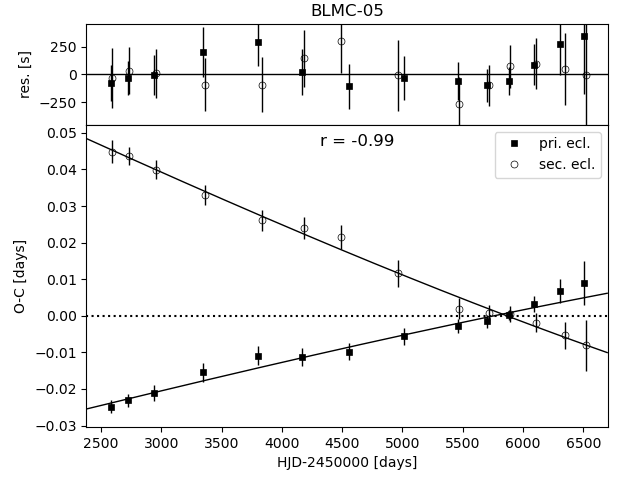}
      \includegraphics[width=0.32\textwidth]{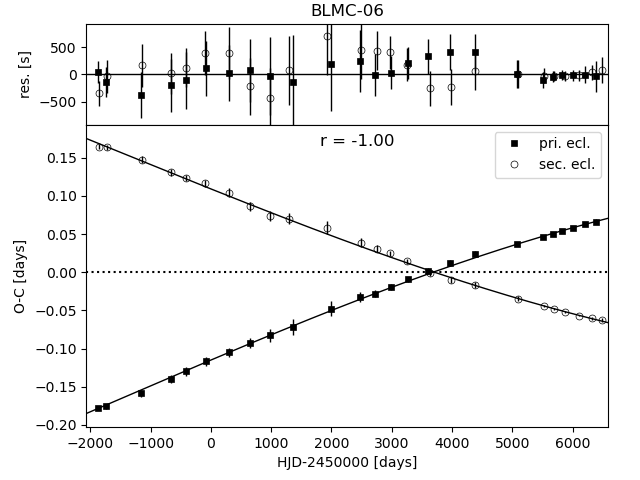}
    \end{center}
    \caption{O-C diagrams with residuals for BLMC-04, BLMC-05, and BLMC-06. Phase shifts of primary and secondary eclipses from a model with constant $\omega$ obtained for OGLE-III data only, calculated along all available data, are shown. As expected for apsidal motion, the shifts are anti-correlated (r close to -1). Small systematic and correlated (for both eclipses) deviations seen for BLMC-04 are probably due to an influence of a fourth component on a wide orbit}. 
    \label{fig:oc_aps}
\end{figure*}

The gravitational interaction of binary system components and stellar rotation make the star shape deviate from a sphere. For a binary orbit, this also causes the rotation of the line of apsides, called apsidal motion, which produces a cyclic variation of the eclipse's position and the light curve shape in general. In a binary model, this variation is represented by a first derivative of the argument of periastron, $\dot{\omega}$.
We detected a significant apsidal motion in the modeling of light and RV curves for all analyzed systems. This effect can be clearly seen in the O-C (observed minus calculated) diagrams (see Fig.~\ref{fig:oc_aps}), which show the phase shifts for the primary and secondary eclipses. These shifts were calculated as in \citet{pilecki:2021} (see also Paper III) but separately for phases around each eclipse (i.e, not for the whole cycle). 
As expected for apsidal motion, shifts of primary and secondary eclipses are anti-correlated for our systems, with a correlation parameter $r$ close to $-1$.  However, correlated systematic deviations from the fit can be seen in the residuals for BLMC-04, where the position of eclipses is affected by another phenomenon described in the following subsection. For BLMC-05, there is no clear systematic deviation, while for BLMC-06, a slight systematic deviation between 2000 and 4500 days can be seen. 

Depending on the system, the analysis had to consider additional techniques or phenomena to obtain a proper model. In the following subsections, we describe the part of the analysis that was particular for a given binary system.

\subsection{BLMC-04 -- Multiplicity}
\begin{figure}
    \begin{center}
      \includegraphics[width=0.49\textwidth]{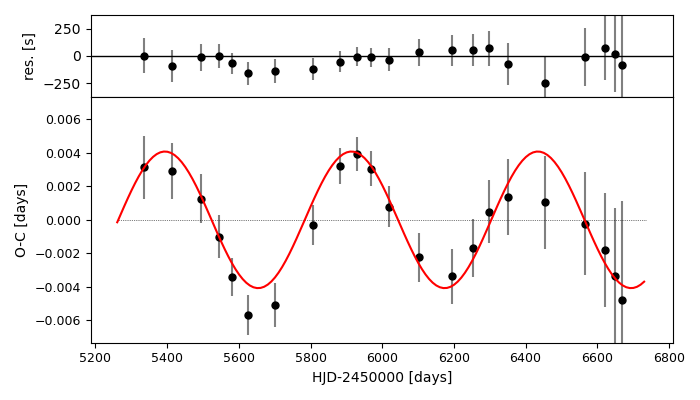}
    \end{center}
    \caption{O-C diagram for BLMC-04 using only the OGLE-IV Ic-band data. Phase shifts for different time bins with respect to the model with constant $\omega$ are shown with the LTTE fit (red line). The orbital period related to the extra component is about 520 days.
    }
    \label{fig:oc_3rdBody}
\end{figure}

During the analysis of BLMC-04, we found that the deviations of RVs from the model are high and similar for both components. Moreover, in the light curve residuals, an extra scatter was present around eclipses, similar to what we found for the multiple system BLMC-03 (see Fig.~5 in Paper III). We consider these two features to be a clear indication of the extra components in the system. Analyzing the residual phase shifts in the O-C diagram, we found two distinct variations depending on the used time resolution. Looking at variations of the order of 1000 days, one can see a possible light-travel time effect (LTTE) due to a companion on a wide eccentric orbit of about 7000-8000 days. This is the variability mentioned in the preceding section, which can be seen in residuals for BLMC-04 in Fig.~\ref{fig:oc_aps}. More data, however, will be needed to confirm it is cyclic and prove the presence of this body.
Using finer time resolution, a shorter period and higher amplitude cyclic variability due to a companion on a tighter orbit can be seen. This variability is the principal cause of the additional RV variation.
From the LTTE fit, we determined the orbital period of the outer orbit related to this (third) body to be $P_{out} \approx$ 520 days (see Fig.~\ref{fig:oc_3rdBody}).
The short-period LTTE can be most clearly seen if we limit the analysis to the OGLE-IV data. These data consist of a large number of measurements in a relatively short time, which minimizes the effect of the apsidal motion and long-period LTTE.

Having this solution, we included the binary orbit around the barycenter of the triple system with the obtained $P_{out}$ in the RV curve analysis. 
This way, we corrected the measured RVs for the presence of the third body, decreasing the scatter from 12.4 $km\,s^{-1}$ to 3.9 $km\,s^{-1}$. We did not consider the effect of the tentative fourth component on the wider outer orbit, but in comparison, it would be insignificant.
We checked the spectra for lines of the extra components but could not detect them, which can be explained by their small contribution to the total system light (only 4\% in the V-band).

A similar check of O-C diagrams was performed for the remaining two systems, but no extra body was detected with confidence in their cases.

\subsection{BLMC-05 -- Spectral Analysis } \label{sec:non-lte}
The color-temperature calibrations do not work well for O-type systems. Therefore, the temperature of the components of BLMC-05 must be determined using non-LTE spectroscopic analysis.

We proceed similarly as in Papers II and III. Using the {\tt FASTWIND} non-LTE model atmosphere code \citep{puls:2005}, we calculate normalized flux spectra for the primary and secondary stars using the stellar gravities derived from the radii and masses obtained from the fit of the light and radial velocity curves. With the stellar effective temperatures of the two components treated as a free parameter, we produce two sequences of models at fixed gravity for the primary and secondary. We then combine the primary and secondary spectra into a composite spectrum by accounting for both components' V-band luminosity contributions and phase-dependent radial velocity-induced spectral shifts. The computed spectra are then compared with our UVES and MIKE observed spectra. We use  UVES spectra at phases 0.25 (3 spectra combined), 0.6 (3 spectra), and 0.65 (2 spectra) and MIKE spectra at phases 0.2 (2 spectra) and 0.8 (3 spectra). We use the co-added spectra to obtain an improved signal-to-noise ratio.

\begin{figure}
    \begin{center}
        \includegraphics[width=0.45\textwidth]{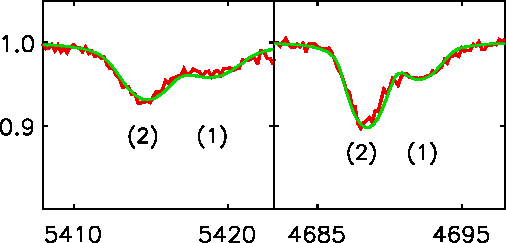} \\
        \includegraphics[width=0.45\textwidth]{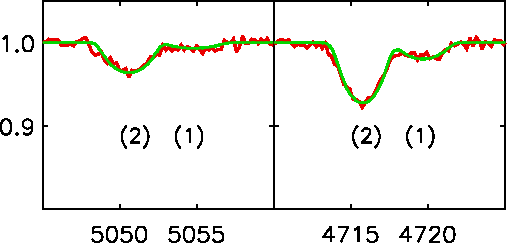}     
    \end{center}
    \caption{Example fit of HeII lines (top panels) of the UVES spectrum at phase 0.6. Left: HeII 5411, right: HeII 4686. The line contributions from the primary and secondary stars are indicated by (1) and (2), respectively. Similarly, fits of HeI lines are shown in the bottom panels. Left: HeI 5048, right: HeI 4713.}
    \label{fig:fig_ab}
\end{figure}

The effective temperatures of the primary and secondary are then obtained from a fit of all HeI and HeII lines available in our spectra using the $\chi^2$ values as described in detail in Papers II and III. We obtain $T1=36000\pm500K$ and $T2=31500\pm500 K$ together with a helium abundance of $N(He)/N(H)=0.1\pm0.01$ and a microturbulence of 10 $km\,s^{-1}$. As an example of the quality of the results, in  Fig.~\ref{fig:fig_ab}, we show fits for two HeII and two HeI lines.

The observed V-band luminosity ratio, derived from the binary modeling, is $L1/L2=3.18\pm0.06$. We can use the model SEDs of our final fit for the primary and secondary, together with the V-band filter function, to calculate the corresponding ratio of our model fit. We obtain $L1/L2=3.23$ in agreement with the observations.

\begin{figure*}
    \begin{center}
        \includegraphics[width=0.32\textwidth]{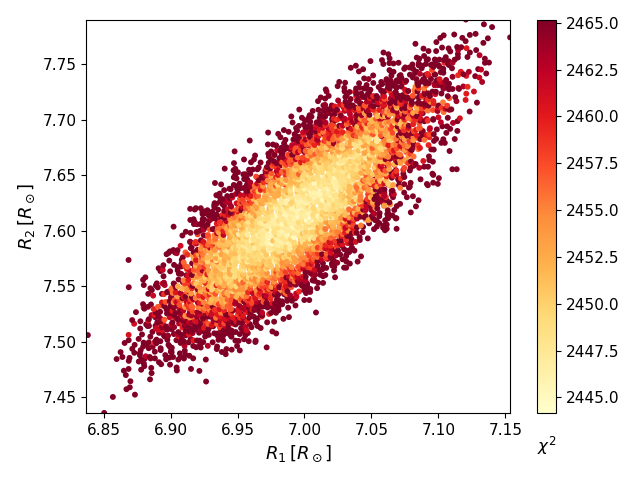} 
        \includegraphics[width=0.32\textwidth]{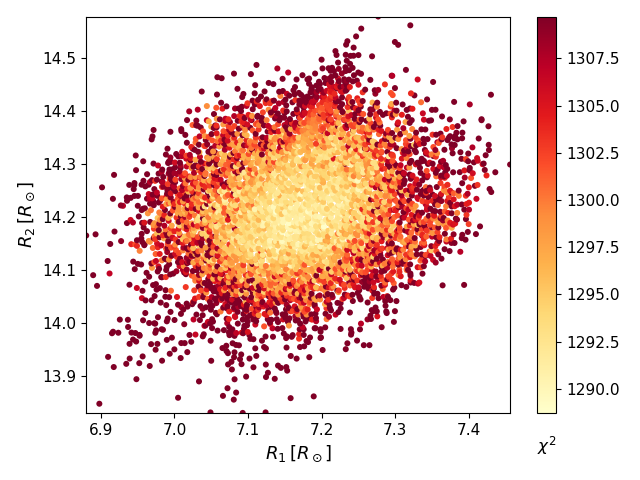} 
        \includegraphics[width=0.32\textwidth]{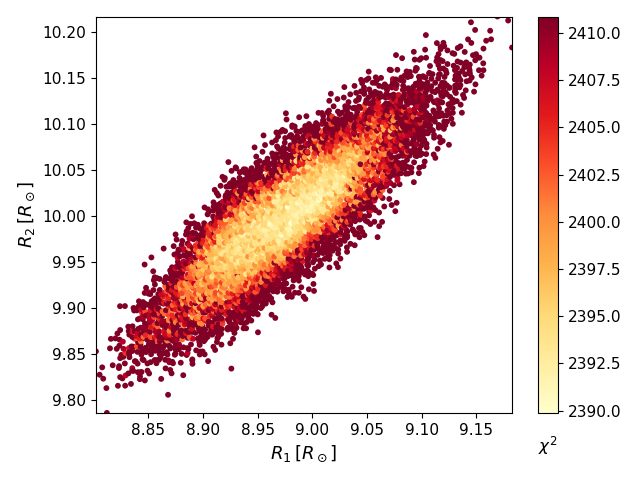} 
    \end{center}
    \caption{Dependence of the $\chi^2$ on the radius of the primary and secondary for the three analyzed systems BLMC-04, BLMC-05, BLMC-06 (from left to right). The presented parameter space is divided into bins; each point is the best solution inside a given bin.}
    \label{fig:radcorr}
\end{figure*}

An interesting feature of this binary is that it exhibits a wide total eclipse because of the high inclination and a large difference in the radii of the components. This makes the analysis more straightforward and reliable as it removes the degeneracy between the radii, as can be seen in Fig.~\ref{fig:radcorr} (middle panel).

\subsection{BLMC-06 -- Period Change }

\begin{figure}
    \begin{center}
        \includegraphics[width=0.48\textwidth]{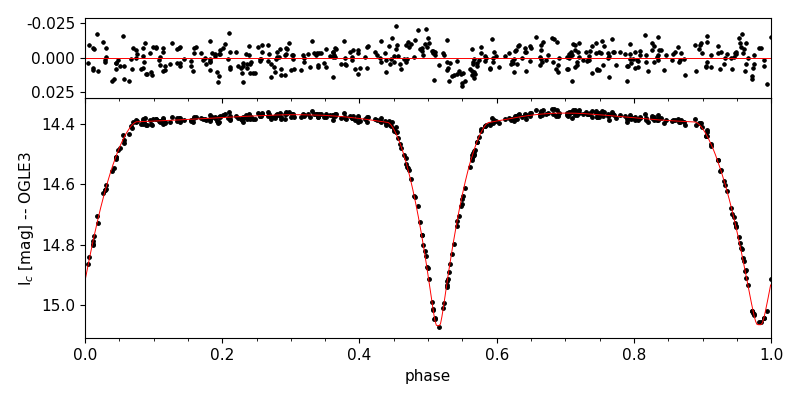} \\
        \includegraphics[width=0.48\textwidth]{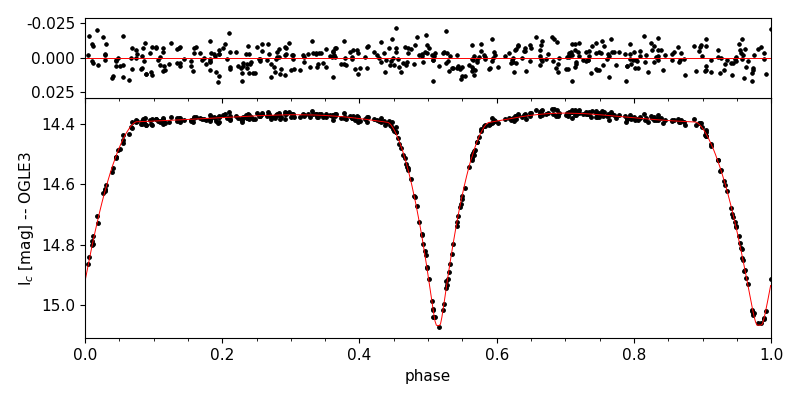} \\
    \end{center}
    \caption{OGLE-III $I_\mathrm{C}$-band light curve models for BLMC-06 taking (bottom) and not taking (top) into account the period change. The best model fits are shown as a red line. When period change is included, the $\chi^2$ for the presented light curve decreases significantly (from 500 to 453). The remaining deviation indicates the period change may not be linear. Both light curves were corrected for apsidal motion. }
    \label{fig:lcs_scat}
\end{figure}

During the preliminary light curve analysis, which used only OGLE-III data, we obtained a good fit for this system with no significant residuals.
However, when all data sets were included, systematic deviations from the model were detected, mainly in the OGLE-III data in eclipses, as seen in the top panel of Fig.~\ref{fig:lcs_scat}. The orbital period determined for all data sets was also significantly different from that obtained in the preliminary solution, so we concluded that the system's orbital period changes and included the period change ($\dot{P}$) in the fitting process. A significant period decrease rate was detected, and systematic deviations mostly disappeared from residuals (bottom panel). The slight deviation that remained indicates that the period change is nonlinear. A possible explanation would be a close passage of a long-orbit unseen companion, but no third light was detected for this system. Looking at the O-C residuals in Fig.~\ref{fig:oc_aps} gives no clue as no trend can be seen there, with apparently only the central part of the data being shifted in phase. Hence, the subject is open for further investigation.

\section{Results}\label{sec:results}

\begin{figure*}
    \begin{center}
        \includegraphics[width=0.3\textwidth]{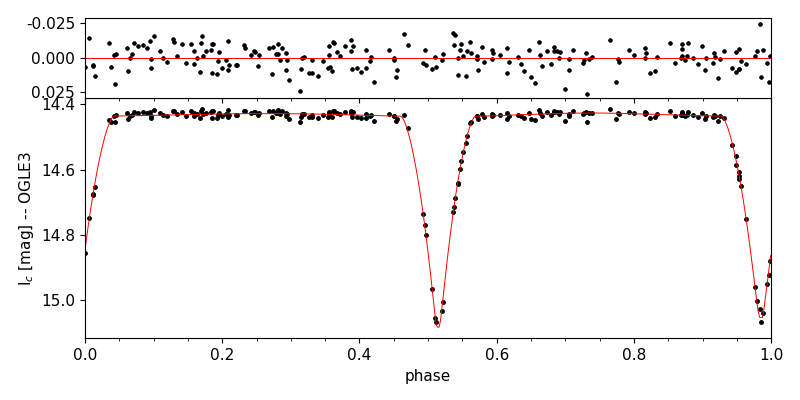} 
        \includegraphics[width=0.3\textwidth]{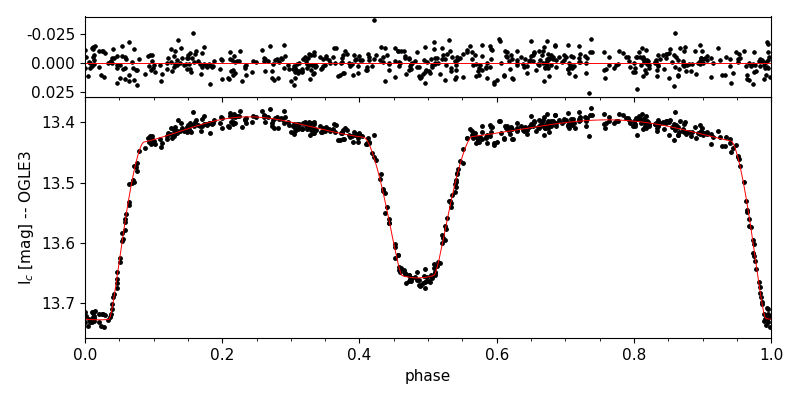} 
        \includegraphics[width=0.3\textwidth]{lc_Io3_blmc06corr} \\
        
        \includegraphics[width=0.3\textwidth]{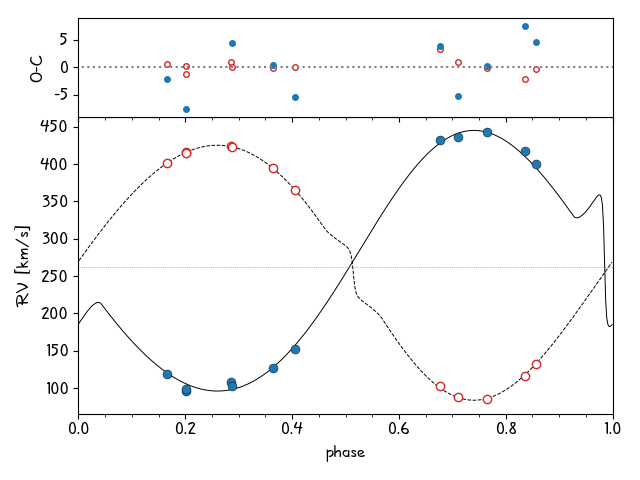} 
        \includegraphics[width=0.3\textwidth]{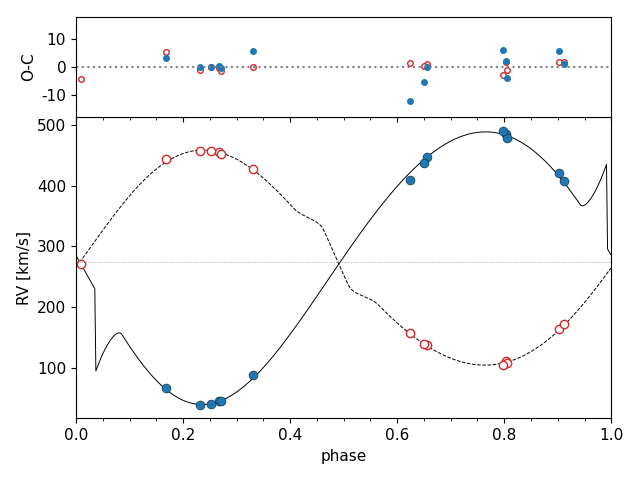}
        \includegraphics[width=0.3\textwidth]{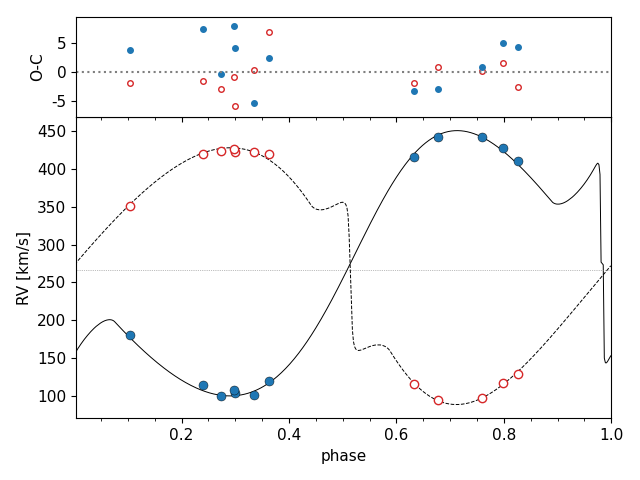} \\
    \end{center}
    \caption{BLMC-04 (left panel), BLMC-05 (center panel) and BLMC-06 (right panel). Solutions for the Ic-band light curves (top panels) and RV curves (bottom panels) are shown.}
    \label{fig:lc_rvc_all}
\end{figure*} 

\begin{figure}
    \begin{center}
        \includegraphics[width=0.5\textwidth]{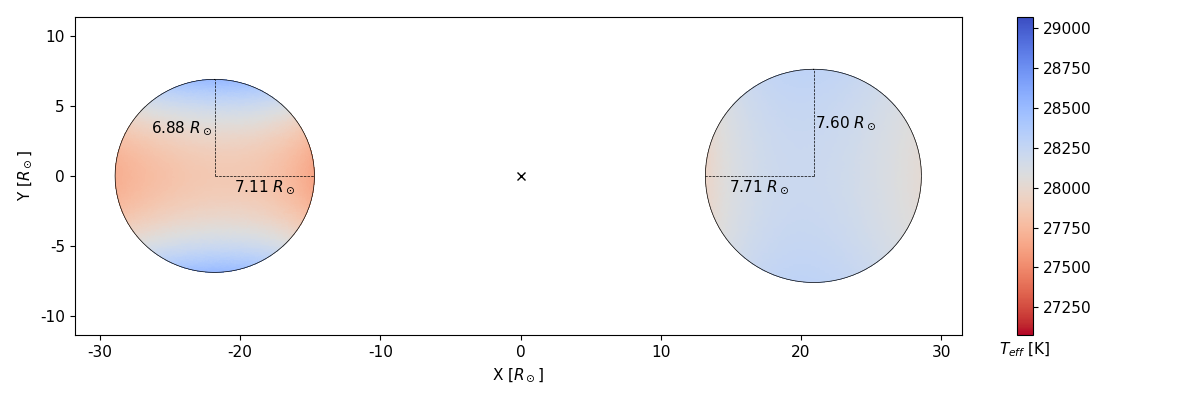} \\
        \includegraphics[width=0.5\textwidth]{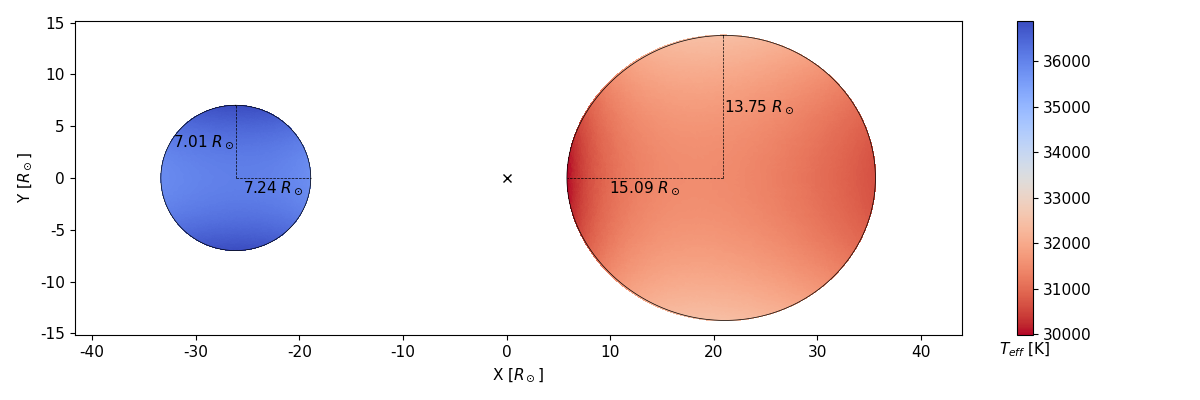} \\
        \includegraphics[width=0.5\textwidth]{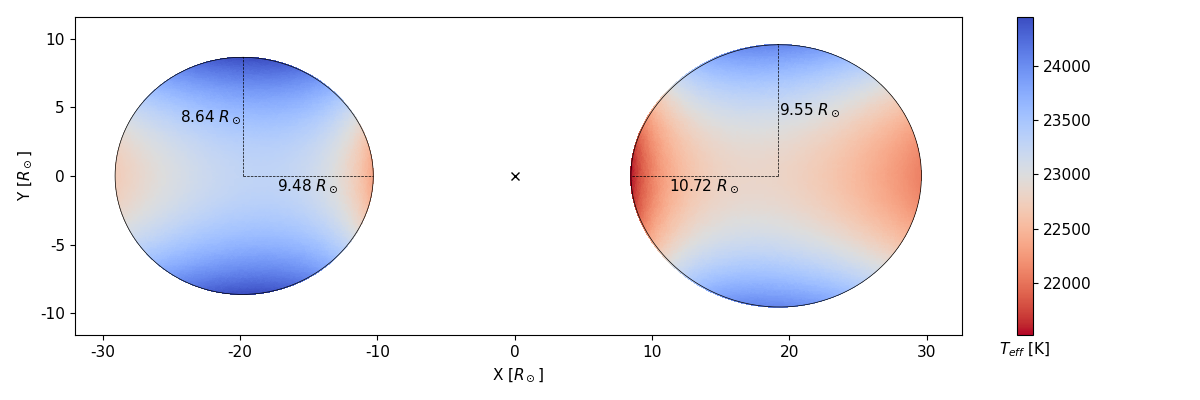} 
    \end{center}
    \caption{Side view for the analyzed systems, from top to bottom: BLMC-04, BLMC-05, BLMC-06. The barycenter is marked with an x, and the shortest (polar) and longest (towards the barycenter) radii are given. The temperature distribution over the star's surface is color-coded. For BLMC-04 the temperature difference between the components is small ($\Delta T \sim 200$K) and for easier comparison we kept the same temperature span of 2000 K in the color bar as for the other B-type system, BLMC-06. The primary of the O-type BLMC-05 is much hotter and smaller than the secondary and the temperature span is much larger.
    }
    \label{fig:sideview}
\end{figure}

Considering all the effects described above, we obtained the final binary model for the analyzed systems. In Fig.~\ref{fig:lc_rvc_all}, we present the Ic-band light and orbital RV curves corrected for apsidal motion; the best models are shown for a specific reference time, while data points are shifted to make up for the model difference between the reference time and the time of observation.
In Fig.~\ref{fig:sideview}, we show a side view of the systems (at quadrature, $i=90^{\circ}$) with star sizes to scale with the separation. One can also see the shape of the stars and the temperature distribution over the surface obtained from the model.

\begin{table*}
\scriptsize
    \begin{center}
    \caption{Physical parameters and other properties of the systems}    \label{tab:results}
    \begin{tabular}{l cc cc cc}
    \hline \hline
    System                         &   \multicolumn{2}{c}{ BLMC-04}         &  \multicolumn{2}{c}{ BLMC-05}          &  \multicolumn{2}{c}{ BLMC-06}  \\
    Parameter [unit]               &     Primary       & Secondary          &     Primary       & Secondary          &     Primary      & Secondary \\
    \hline            
    mass [$M_\odot$]               & 13.05 $\pm$ 0.12  & 13.54 $\pm$ 0.21   & 17.57 $\pm$ 0.15  &  22.1  $\pm$ 0.3   & 11.83 $\pm$ 0.12   & 12.25 $\pm$ 0.14  \\ 
    radius [$R_\odot$]             &  6.99 $\pm$ 0.03  &  7.61 $\pm$ 0.03   &  7.17 $\pm$ 0.07  &  14.21 $\pm$ 0.07  &  8.98 $\pm$ 0.03   & 10.00 $\pm$ 0.04  \\
    $\log g$ [cgs]                 & 3.8642$\pm$ 0.0021 & 3.807 $\pm$ 0.004 & 3.972 $\pm$ 0.008 &  3.477 $\pm$ 0.003 & 3.6046$\pm$ 0.0025 & 3.5259$\pm$ 0.0022 \\ 
    temperature [K]                & 27800 $\pm$ 1500  & 28000 $\pm$ 1500   & 36000 $\pm$ 500   &  31500 $\pm$ 500   & 23500 $\pm$ 500   & 23000 $\pm$ 500  \\ 
    $(T_2 / T_1)_{phot}$           & \multicolumn{2}{c}{$ 1.0067 \pm 0.0012$}   & \multicolumn{2}{c}{$0.8668 \pm 0.0023$}  & \multicolumn{2}{c}{$ 0.9794  \pm 0.0006 $}   \\
    $\log L$ [$L_\odot$]           & 4.434 $\pm$ 0.004 & 4.517 $\pm$ 0.004 & 4.907 $\pm$ 0.010 & 5.253 $\pm$ 0.004  &  4.344 $\pm$ 0.004 & 4.402 $\pm$ 0.003  \\
    $j_{21,V} = F_{2}/F_{1} (V)$   & \multicolumn{2}{c}{1.0159 $\pm$ 0.0017} & \multicolumn{2}{c}{ 0.806 $\pm$ 0.003} & \multicolumn{2}{c}{0.9725 $\pm$ 0.0009 }  \\
    $v_{rot}$ [$km\,s^{-1}$]       &         134      &    25            &    145        &  135 & 150 & 152  \\
    oblateness                     &    0.03              &    0.01      & 0.03  &  0.09 & 0.09  & 0.11   \\
    reference time, $T_0$ [days]   & \multicolumn{2}{c}{ 2455855.66132(17)} &  \multicolumn{2}{c}{ 2453577.3285(6) } & \multicolumn{2}{c}{2453571.03213(19) }   \\
    orbital period, $P$ [days]     & \multicolumn{2}{c}{6.2290957(4) } & \multicolumn{2}{c}{ 5.9468115(22)} &  \multicolumn{2}{c}{ 5.7259370(4) }     \\
    semi-major axis [$R_\odot$]    & \multicolumn{2}{c}{ 42.52 $\pm$ 0.17 } & \multicolumn{2}{c}{ 47.10 $\pm$ 0.16}   &  \multicolumn{2}{c}{38.89 $\pm$ 0.13}   \\
    orbital inclination, $i$       & \multicolumn{2}{c}{ 89.88 $\pm$ 0.10 } & \multicolumn{2}{c}{ 87.7 $\pm$ 0.6}   &  \multicolumn{2}{c}{ 89.16 $\pm$ 0.15}     \\
    eccentricity, $e$              & \multicolumn{2}{c}{ 0.0554  $\pm$ 0.0006 } & \multicolumn{2}{c}{0.0635 $\pm$ 0.0011} & \multicolumn{2}{c}{ 0.1338 $\pm$ 0.0005 }       \\
    argument of periastron, $\omega$ [rad]    & \multicolumn{2}{c}{ 5.715 $\pm$ 0.016}  & \multicolumn{2}{c}{ 2.338 $\pm$ 0.016} & \multicolumn{2}{c}{ 5.0948 $\pm$ 0.0019 }  \\
    apsidal motion, $\dot{\omega}$ [rad/d]    & \multicolumn{2}{c}{ 0.0000576(15) }     & \multicolumn{2}{c}{ 0.000123(6)}       & \multicolumn{2}{c}{0.0001234(6) }  \\
    systemic velocity, $\gamma$ [km s$^{-1}$] & \multicolumn{2}{c}{ 284.0 $\pm$ 0.3 } & \multicolumn{2}{c}{274.0 $\pm$ 0.5}    & \multicolumn{2}{c}{ 266.6 $\pm$ 0.4 }  \\
    mass ratio, $q = m_2/m_1$                 & \multicolumn{2}{c}{ 1.038 $\pm$ 0.008 } & \multicolumn{2}{c}{1.258 $\pm$ 0.007}  & \multicolumn{2}{c}{ 1.033 $\pm$ 0.005 }  \\
    \hline
    \end{tabular}
    \end{center}
    \tablecomments{For the reference time, orbital period, and the argument of periastron errors in the last digits are quoted in parenthesis.}
\end{table*}

In Table~\ref{tab:results}, we provide the properties of the analyzed systems and their components.
As the stars are not perfectly spherical, the radii of the stars in the table correspond to the radii of spheres that would have the same volume as the stars. As a measure of oblateness, a value $(R_{max} - R_{min})/R_{max}$ is given.
In the study, we use the nominal solar constants following the IAU 2015 Resolution B3 \citep{prsa2016}.

\subsection{The Six LMC DEBs}
With this study, we doubled the number of objects characterized in the project, which allows us to conduct various comparisons. 
The six systems analyzed to date form a diverse collection of early-type binaries that can be found in other galaxies in the Local Group, where the SBCR method can be applied. In the analyzed sample, there is one system with a total eclipse, three with a significant third light, two where the inclusion of a third body was necessary for the modeling of light and RV curves, one with a significant period change, and one with a practically circularized orbit (having eccentricity $e=0.0006 \pm 0.0003$). A summary of these properties is provided in Table~\ref{tab:orbits}. This table also includes $\dot{\omega}$, the apsidal motion period U, and the third light contribution in the V band for the six systems. We also indicate whether the orbital motion due to additional components was detected in the O-C analysis or as systematic shifts in radial velocities.

\begin{deluxetable*}{lccccccccc}
    \tabletypesize{\footnotesize}
	\tablecaption{Selected properties of systems in our sample} \label{tab:orbits}
         \tablehead{
                \colhead{Our ID} &  \colhead{Orbital Period} & \colhead{e} & \colhead{$\dot{\omega}$} & \colhead{$U$}  & \colhead{$\dot{P}$}  & \colhead{$3^{rd}$ light (V)}  & \multicolumn{2}{c}{Multiplicity} & \colhead{Comments} \\
                \colhead{  }     &  \colhead{[days]}         & \colhead{}    & \colhead{[rad/day]} & \colhead{[yr]} & \colhead{}              & \colhead{[mag]}          & \colhead{3rd} & \colhead{4th}     & \colhead{} 
                   }
           \startdata
                BLMC-01  &  5.4139927   & 0.0151(7) & 0.000120(13) & 144(15)                & - & 0.0  & - & - & B-type \\
                BLMC-02  &  4.2707640   & 0.0809(8) & 0.000145(3) & 119(2)                & - & 0.0  & - & - & O-type\\
                BLMC-03  &  3.2254367   & 0.0006(3) & -0.005(4)    & ---\tablenotemark{a} & - & 0.18 & O-C/RV & O-C & O-type  \\
            \hline                                                                                   
                BLMC-04  &  6.229096   & 0.0554(6) & 0.0000576(15) & 298(7)               & - & 0.04 & O-C/RV & O-C\tablenotemark{b} & B-type \\
                BLMC-05  &  5.946812   & 0.0635(11)& 0.000123(6)  & 140(6)              & -  & 0.06 & - & - & O-type, TECL    \\
                BLMC-06  &  5.7259370  & 0.1338(5) & 0.0001230(6) & 139.8(7)    & -5.0$\times$$10^{-9}$  & 0.0  & - & - & B-type \\
          \enddata
\tablecomments{Three first systems were published in Papers I, II, and III. Results for the three last systems come from this work. In the multiplicity column, we show by what means the 3rd and 4th components were detected, either in the O-C analysis or in RV residuals. TECL means the system shows a clear total eclipse.
}
\tablenotetext{a}{The eccentricity is too low to determine the apsidal motion period (U) for this system with sufficient precision.}
\tablenotetext{b}{With clear LTTE signature, but data coverage is insufficient to prove cyclic behavior.}
\end{deluxetable*}

\subsection{Internal Structure}
The secular change in the position of the periastron is a consequence of distortions, such as tides and rotational flattening, that affect the structure of the components in close eccentric binaries. These distortions can be described as a function of the internal structure constant (ISC), $k_2$ \citep{kopal:1978}, related to the density distribution inside a star. Although we cannot measure this parameter directly for each component, we can calculate a specific weighted mean for both stars from the measured apsidal motion's period. The mean value of the ISC is given by

$$ \overline{k}_{2,obs} = \frac{1}{c_{21} + c_{22}} \frac{\dot{\omega}}{ 360}, $$

\noindent where $c_{21}$ and $c_{22}$ are functions of the orbital eccentricity, fractional radii, the masses of the components, and the ratio between the rotational velocity of individual stars and orbital angular velocity \citep[see][]{kopal:1978}.

Having accurate absolute sizes of the components of our six binary systems, we can investigate the stellar interior, comparing the theoretical and observed values of the ISC.
In the calculations, we assumed that the rotational axes are aligned with the orbital axis.
The observed apsidal motion rate has two independent components: the Newtonian (classical) term, due to the non-spherical shape of both stars, and the relativistic term, due to general relativity effects. To be able to compare the theoretical and observed value, the relativistic term, $\dot{\omega}_{rel}$, must first be subtracted from the latter.  This term can be calculated using the following formula \citep{gimenez:1985}:

$$ \dot{\omega}_{rel} = 5.45 \,x\, 10^{-4} \frac{1}{1-e^2} \left( \frac{M_1+M_2}{P} \right)^{2/3},$$

\noindent where $M_i$ denotes the individual masses of the components in solar units, $e$ is eccentricity, and $P$ is the orbital period in days. The values of $\dot{\omega}_{rel}$ and $\overline{k}_{2,obs}$ for each system, are listed in Table~\ref{tab:interna_struct}.

For the theoretical values of ISC, well-calibrated evolutionary models are necessary. We used theoretical models focused on internal structure computed by \cite{claret:2019}, with the $k_{2,theo}$ term provided along the evolutionary tracks. To obtain its value for the components of our systems, we looked for the closest model in the parameter space of $\log T$, $\log g$, and mass for two of the available metallicities, Z=0.0042 and Z=0.0134. Then, we interpolated those values at the approximate average LMC metallicity of Z=0.008. Finally, the mean values, $\overline{k}_{2,theo}$, were calculated for each system. They are included in Table~\ref{tab:interna_struct}.
The comparison between theoretical and observational values shows that the agreement for O-type binaries is very good, while the difference is significant for B-type ones. It is, however, unclear if this is a coincidence or if there is a reason for that.  In the case of BLMC-04, one possible cause is the presence of additional components on relatively close orbits, but no such presence was detected for others. To check if the model resolution in mass could cause this effect, we chose the second closest track in mass but found very similar $\overline{k}_{2,theo}$ values.
In further study, a more complex approach that includes evolutionary models calculated individually for each system will be essential to understand when and why the observational and theoretical values differ.

\begin{deluxetable*}{lccccc}
    \tabletypesize{\footnotesize}
	\tablecaption{Internal structure} \label{tab:interna_struct}
         \tablehead{
                \colhead{System ID}  & \colhead{$\dot{\omega}$}   & \colhead{$\dot{\omega}_{rel}$}  & \colhead{$\dot{\omega}_{rel}/\dot{\omega}$} & \colhead{log$\,k_{2,obs}$}  & \colhead{log$\,k_{2,theo}$}  \\
                \colhead{  }   & \colhead{[deg/cycle]}    & \colhead{[deg/cycle]} & \colhead{(\%)}   & \colhead{} & \colhead{}
                 }
           \startdata
                BLMC-01 & 0.037(4) & 0.001616(6) & 0.044(5)      &  -2.86(5)    & -2.602 \\ 
                BLMC-02 & 0.0354(7) & 0.002380(10) & 0.0672(13)  & -2.028(2)    &  -2.038 \\ 
                BLMC-03 & --- & ---  &  --- &  --- & -2.111 \\
                BLMC-04 & 0.0206(5) & 0.001442(9) &  0.0700(17)   &  -2.056(17) & -2.255  \\
                BLMC-05 & 0.0417(19) & 0.001938(10) & 0.0464(21)  &  -2.548(25) & -2.573  \\
                BLMC-06 & 0.04037(20) & 0.001446(7)  &  0.03582(25) & -2.584(11) & -2.486 \\          
            \enddata
\end{deluxetable*}

\subsection{Mass-luminosity relation}
\label{sec:mlr}
When the luminosity of main sequence stars (i.e., stars that are fusing hydrogen into helium in their cores) is plotted against their masses, we observe the stars to form a narrow band called the mass‐luminosity (M-L) relation. This relationship is considered one of the most fundamental descriptions of stellar properties. 

\cite{gonzalez:2005} determined for the first time a well-constrained empirical mass-luminosity relation for late O and early B type stars in the LMC (see their Fig. 15). For that, they used eight DEBs analyzed by them, supplemented with similar systems studied by other authors. The latter are three binaries, including HV 2274, used in an earlier distance-scale project based on early-type systems \citep{guinan:1998b, ribas:2002, fitzpatrick:2002} and five high-mass binaries in 30 Doradus nebula \citep{massey:2002,massey:2012}. 

\begin{figure}
    \begin{center}
        \includegraphics[width=0.45\textwidth]{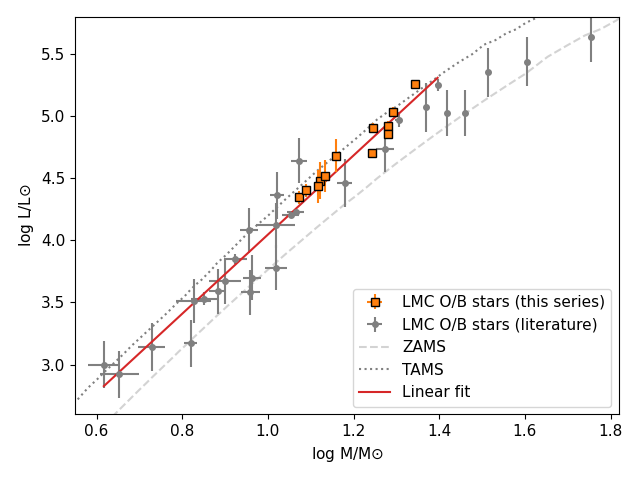} \\
        \includegraphics[width=0.45\textwidth]{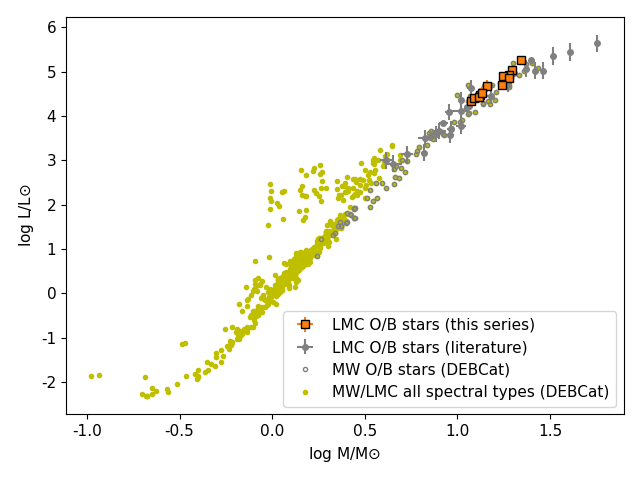}
    \end{center}
\caption{\textit{Top panel:} Empirical M-L diagram showing the relationship between mass and luminosity for O- and B-type components of the LMC DEBs. Theoretical limits for the main sequence (values for ZAMS and TAMS) at LMC metallicity are overplotted. 
The components of our systems (orange) lie systematically closer to TAMS. Error bars for them are mainly smaller than the point size, especially for the masses.
A linear fit to the data in a range of 4 M$_\odot$ to 25 M$_\odot$ is also shown. \textit{Bottom panel:} A similar diagram extended with the MW O- and B-type systems and all (MW and LMC) systems of spectral types A and later from DEBCat.}
\label{MLrel}
\end{figure}

We extended this sample with components of four new DEBs analyzed in this and previous papers of the series (Paper I, II) and used new results for one of the binaries in 30 Doradus (Paper III) and for HV 2274 (this work). The M-L relation for such an updated sample is presented in the top panel of Fig.~\ref{MLrel}. Our measurements contribute significantly to the mass range from 11 to 23 M$_\odot$.
The statistics at the high-mass end are still low, but the available data coincide with a slight change in the curvature of theoretical limits for main sequence stars.
The M-L relations for zero-age main sequence (ZAMS) and terminal age main sequence (TAMS) shown in the plot come from evolutionary tracks from MESA Isochrones \& Stellar Tracks models \citep[MIST,][]{dotter:2016,choi:2016} for metallicity close to the LMC, [Fe/H] = -0.5 dex. We checked that using a metallicity of -0.25 dex does not change these limits significantly, so we did not interpolate tracks to the precise LMC value. Also, the difference between models taking and not taking rotation into account is negligible.

In Fig.~\ref{MLrel}, we also show a linear fit to the data in the range of 4 M$_\odot$ to 25 M$_\odot$ using the orthogonal distance regression method and uncertainties in both masses and radii. Because of high precision, our data points dominate in the upper part of the relation. We want to highlight the importance of combining binary modeling with spectroscopic analysis to obtain precise effective temperatures and luminosities. The error bars are smaller than the point size for our DEB components with such an analysis. The fitted relation is:

$$ \log L/L_\odot = 3.18  \pm 0.12 \log M/M_\odot + 0.86 \pm 0.14 $$

In the bottom panel, we show a similar plot extended with components of late and early-type detached eclipsing binaries from the DEBCat catalog \citep{southworth:2015}. These binaries have physical parameters determined with precision better than 2\%. Currently, DEBCat provides data for 334 systems. This includes 47 O- and B-type binaries, of which 45 are Galactic and two from the LMC. However, the latter are our BLMC-01 and BLMC-02, the parameters of which in DEBCat come from Papers I and II.  Note that we have not cleaned the DEBCat sample of evolved stars, which are more luminous for the same mass and, therefore, lie above the relation.

\subsection{Mass-radius relation}
\begin{figure}
    \begin{center}
        \includegraphics[width=0.45\textwidth]{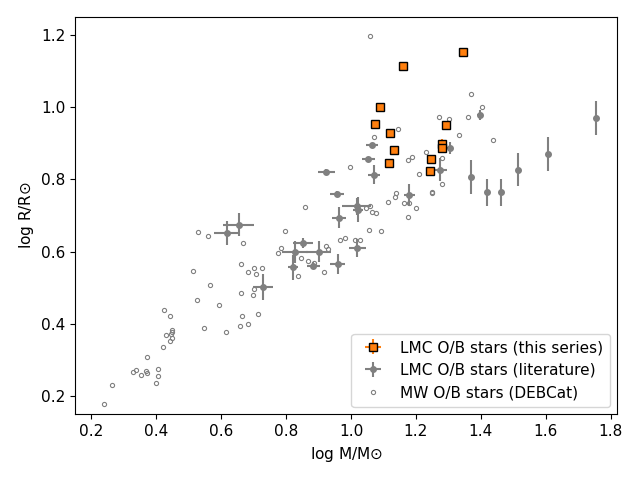} \\
        \includegraphics[width=0.45\textwidth]{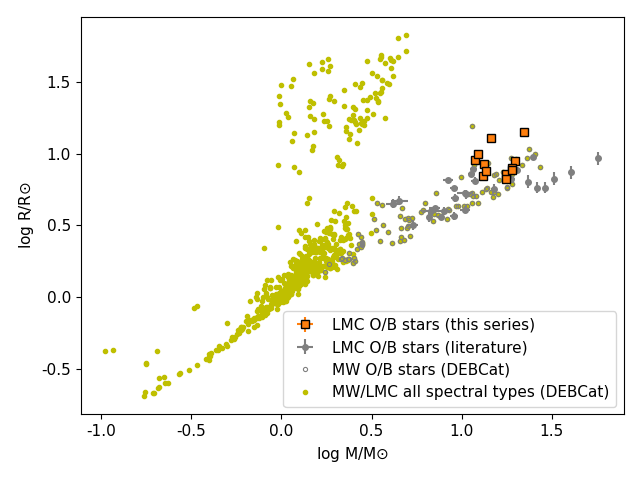}
    \end{center}
\caption{\textit{Top panel:} The position of our stars (BLMCs) in the M-R diagram. Components of the O- and B-type DEBs from DEBCat are included for comparison. Error bars for our measurements are smaller than the point size.
The radii of the components of our systems are generally larger than those of other stars with similar masses, indicating that they are more evolutionary advanced on the main sequence.
\textit{Bottom panel:} Same diagram but also including systems of spectral types A and later. }
\label{MRrel}
\end{figure}

The relationship between mass and radius is also quite tight, as shown in the mass-radius (M-R) diagram in Fig.~\ref{MRrel} for the same sample as above. However, evolution can significantly increase stellar radii along the main sequence. This can be seen in the top panel, which shows the early-type binary components from the LMC and MW. The upper boundary is not well defined there, with many points extending toward larger radii. In the bottom panel, we show the same diagram but with the addition of stars of later spectral types, including evolved stars with even larger radii. Our sample contributes significantly to statistics for massive binaries, especially in the late phases of main sequence evolution, as indicated by relatively large radii for their masses.

\subsection{Orbital periods of early-type DEBs}
\begin{figure}
    \begin{center}
        \includegraphics[width=0.45\textwidth]{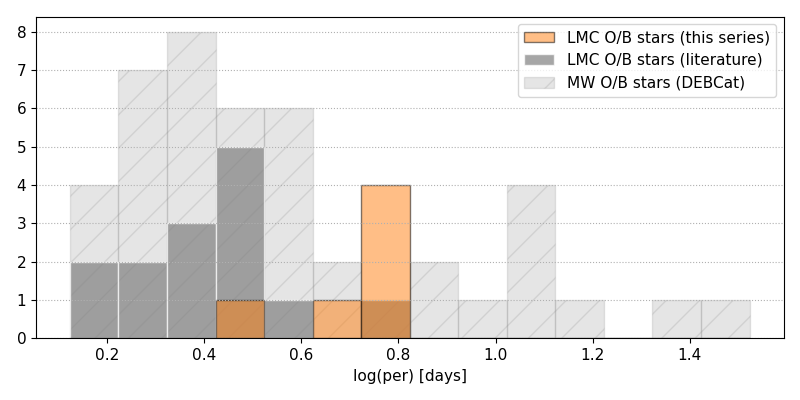}
    \end{center}
\caption{Histogram of orbital periods showing a comparison of our sample with other early-type stars from the LMC and the Galactic ones.} 
\label{perhist}
\end{figure}

In Fig.~\ref{perhist}, we show the period distribution for the 45 Galactic binaries composed of O and B-type stars from the DEBCat catalog compared with our sample and other LMC early-type systems presented in Section~\ref{sec:mlr} on the mass-luminosity relation. Our systems are mainly in the long-period tail for all such systems and at the long-period end for the LMC systems.

\section{Discussion}\label{sec:discussion}
Along this series of papers, we analyzed six early-type detached eclipsing binary systems that exhibit a variety of configurations and features. To account for that, one has to extend the standard binary modeling of light and radial velocity curves with additional techniques and secure a sufficient number and timespan of observations. Only then can reliable physical parameters and the resulting distances be obtained.

However, we must mention that the systems we analyzed so far had to first pass through the preliminary selection. In addition to the light curve features mentioned earlier, the orbital amplitudes, eccentricity, and rotational velocities of the binary components had to be such that their lines could be separated during a significant part of the orbital cycle; only then is the unambiguous and precise determination of the radial velocities possible. Apart from fulfilling these basic system criteria, two essential features of the analysis must be highlighted: a properly accounted third light that has to be determined directly in all considered bands and the inclusion of the system's orbital motion due to extra components. The first feature is an issue for three of the six systems, with the third light reaching 18\%. Not accounting for this effect may lead to the color of the components being wrongly determined and eventually either wrong distance or surface brightness, depending on whether we use or calibrate the SBCR. Although photometric data are relatively easy to obtain for the calibration sample in the LMC, they are not readily available in sufficient amounts in infrared bands. Moreover, a complete set of data must also be collected for the systems in other galaxies to apply the SBCR method, which will need significant effort.

Another, though connected problem is the presence of the third body or, in general, extra components that affect all datasets because of the light-travel time effect and, additionally, cause extra shifts to measured radial velocities. The impact on the light curves is relatively weak, although sometimes, it may produce significant phase shifts that cannot be ignored. In contrast, the effect on the RVs is generally more substantial. It may lead to wrong masses and radii of the components, while the accuracy of determined radii directly influences all distance determination efforts. A strong effect of this type is present in two of our analyzed systems, and to be detected and correctly accounted for, it was necessary to use a more complex approach to model them. Wide outer orbits strongly affect the light curves and, therefore, can be detected through the O-C analysis. Tighter orbits produce higher additional orbital velocities, systematically shifting measured radial velocities and adding scatter to RV curves. Thus, looking at the O-C diagrams and RV curves is complementary, and both ways should be used to characterize the multiplicity of the system.

With our detailed analysis of six early-type DEBs, we paved the way for the use of such systems as accurate distance indicators.
A separate paper on the calibration of the SBCR for O- and B-type stars is in preparation.

It is interesting to compare our results with those obtained previously for BLMC-06 or HV 2274, the most studied DEB in the LMC. R00 obtained masses of the components 11.4$\pm$0.7 M$_\odot$ and 12.2$\pm$0.7 M$_\odot$, and their radii 9.0$\pm$0.2 R$_\odot$ and 9.9$\pm$0.2 R$_\odot$. These values are in agreement with our results within their error bars. However, our precision is about six times higher. Using our error bars, their best values lie from 1.2 to 5$\sigma$ away, partly inconsistent with our solution.
This shows that reanalyzing binary systems using new data and codes is necessary to use them for high-quality astrophysics.
The improvement in our results came mainly from the more precise RV measurements. The precision estimated by R00 is $\sim$15 $km\,s^{-1}$, while the root-mean-square (or rms) of residuals in our work is only 3.8 $km\,s^{-1}$. 
Another factor is the simultaneous modeling of the RV and light curves and the use of several datasets in different bands spanning about 25 years. This not only helps constrain such parameters as period or third light but also lets us include in the modeling effects such as the apsidal motion or period change. Thus, we could consistently model the system through time and take advantage of looking at it in different orbit configurations.

The study of the apsidal motion of HV 2274 is an interesting topic on its own. The first estimate of this effect was done by W92, who derived U=123$\pm$3 yr 
using the times of minima from their three bands CCD photometry combined with those published by \cite{gaposhkin:1977}. This value was taken as fixed by G98a and R00 in their modeling.
However, W92 remarked on the lack of information concerning the quality of the timings of \cite{gaposhkin:1977} obtained from the plates and noted that the value of U could be better estimated in the following decades with more data. 
In \citeyear{zasche:2013}, \citeauthor{zasche:2013} intended to improve the apsidal motion period of this system. They derived the times of minima from the light curves of W92, MACHO, OGLE-III, and their photometry, and also used the minima times of \cite{gaposhkin:1977}. They found an additional variation superimposed on the apsidal motion, explained as the presence of a third body with a period of 98 years and a minimum mass of 3.4 M$_\odot$. 
The data from \cite{gaposhkin:1977} are crucial for this conclusion, but, as seen in their Fig.~3, the scatter is very high (of the order of 0.05 days), making the presented solution uncertain. Moreover, the timeline of observations is not long enough to ensure this variability is indeed cyclic.
Using the W92 light curves, but not the eclipse timings of \cite{gaposhkin:1977}, we did not find clear evidence of either a third component in the system or a linear period change, as can be seen in Fig.~\ref{fig:oc_aps} (right panel).

\section{Acknowledgments}
The research leading to these results has received funding from the European Research Council (ERC) under the European Union's Horizon 2020 research and innovation program under grant agreement No 951549 (project UniverScale). 
RPK has been supported by the Munich Excellence Cluster Origins funded by the Deutsche Forschungsgemeinschaft (DFG, German Research Foundation) under Germany's Excellence Strategy EXC-2094 390783311. B.P. acknowledges support from the Polish National Science Center grant SONATA BIS 2020/38/E/ST9/00486.
We also acknowledge support from the Polish Ministry of Science and Higher Education grant DIR-WSIB.92.2.2024. The research was based on data collected under the Polish-French Marie Skłodowska-Curie and Pierre Curie Science Prize awarded by the Foundation for Polish Science.
This research is based on observations collected at the European Southern Observatory under the following ESO programs: 096.D-0170 (B) (PI W. Gieren), 097.D-0400(A), 098.D-0263(A), 0100.D-0339(A), and 0102.D-0469(A,B)(PI G. Pietrzyński), 102.D-0590(B) (PI B. Pilecki). 

This research has made use of NASA's Astrophysics Data System Service.

\vspace{5mm}
\facilities{VLT:Kueyen (UVES), Magellan:Clay (MIKE)}

\software{
\texttt{ESO Reflex} \citep[][\url{http://www.eso.org/sci/software/esoreflex/}]{freudling:2013}, \\
\texttt{PHOEBE} \citep[][\url{http://phoebe-project.org/1.0}]{prsa:2005},\\
\texttt{RaveSpan} \citep[][\url{https://users.camk.edu.pl/pilecki/ravespan/index.php}]{pilecki:2017}, \\
\texttt{FASTWIND} \citep[][]{puls:2005, rivero:2012}}

\bibliographystyle{aasjournal}
\bibliography{taormina2024b_paper4}

\appendix{
\section{Notes on individual systems}\label{sec:app}
In the main body of the text, we provided only the most relevant information about the analyzed systems. However, for completeness, we present a short literature review for each of them below. 

To avoid distracting the reader, we skipped a background description for the analyzed systems in the main body of this paper. However, as a complement to our study, we present a short literature review of them below.

\subsection{ BLMC-04 (OGLE LMC-ECL-17660) } 
This target is located in a dense field in the northeast part of the LMC. There are entries for this object in several photometric catalogs, but it has not been part of any study in the past. 
As mentioned in Section~\ref{sec:data}, we found this system in the MACHO survey under ID 82.8162.24, in EROS2 under ID lm0020n6294, and in different campaigns of the OGLE project as LMC168.6 22300, LMC517.05.033821 \citep{tisserand:2007, alcock:1997, udalski:2015}. In the OGLE catalog of eclipsing binaries, it was classified as a non-contact binary, and its period was determined as P=6.2290904 days} \citep{pawlak:2016}. Apart from that, no other relevant information was available in the literature.

\subsection{ BLMC-05 (OGLE LMC-ECL-18794) }
This system is also located in the northeast part of the LMC, but in a less dense region than the previous object. It is the brightest system in the sample. It was included in the work of \citet{evans:2015}, where they provided spectral classification from optical spectroscopy of 263 luminous massive stars in the north-eastern region of the LMC to increase the knowledge of the spectral content of the galaxy. They classified the BLMC-05 system as O9.5III + B0. There is, however, no study where the physical parameter would have been determined, and, unfortunately, except fo OGLE, there are no data for BLMC-05 in other multi-epoch surveys (as shown in Table~\ref{table:photomData}).

\subsection{ BLMC-06 (HV 2274) }

This system, identified in the OGLE catalog as LMC-ECL-05764, is the most famous and studied binary system outside the Milky Way galaxy. It was the first used to determine an extra-galactic distance. G98a combined the analysis of the light and radial velocity curves with the fitting of the UV/optical spectrum to determine the distance to the LMC. This method of distance determination is significantly more model-dependent than the SBCR method. Due to that and other problems, mainly with the reddening determination, the values of the distance modulus to the LMC obtained by different authors vary significantly (18.22-18.44 mag) \citep[G98a;][]{udalski:1998,guinan:1998b,nelson:2000,ribas:2000,groenewegen:2001}. For example, because of the lack of details about the fitting procedure of the UV/optical spectrum in G98ab, \cite{groenewegen:2001} re-analyzed it, considering different sets of model parameters, but could not reproduce the earlier results. The story of this system is a clear example of why a well-calibrated empirical method is necessary to measure accurate distances to early-type eclipsing binary systems.

\citet{groenewegen:2001} also provided an interesting and complete historical review of observations and studies related to BLMC-06. Therefore, we do not repeat this here and refer the reader to their work for more details. However, it is essential to highlight that a complete independent modeling of the light and RV curves performed to determine the binary parameters was only done in G98a. Although R00 is also referred to as a source of the system properties, in fact, both solutions are very similar, with a difference mainly in uncertainties.
For example, the masses and radii from R00 are 12.2(7) M$_\odot$, 11.4(7) M$_\odot$, 9.86(24) R$_\odot$, 9.03(24) R$_\odot$, while those from G98a are 12.1(4) M$_\odot$, 11.4(4) M$_\odot$, 9.84(24) R$_\odot$, 9.03(20) R$_\odot$.

In the follow-up studies, other authors were only adjusting the G98a/R00 results by using different variants of the method, new photometry, different reddening determinations, or different model atmospheres to determine the LMC distance with this system. For this reason, we found it essential to reanalyze BLMC-06 using the large set of new high-quality photometric and spectroscopic data that were not available before.

\end{document}